\documentclass[aps,prl,twocolumn,superscriptaddress, letterpaper]{revtex4}
\usepackage{graphicx}
\usepackage{hyperref}

\newcommand{\ket}[1]{|{#1}\rangle}

\begin{document}

\title{Approaching Tsirelson's bound in a photon pair experiment}

\author{Hou~Shun~Poh}
\affiliation{Center~for~Quantum~Technologies, National~University~of~Singapore,
3 Science Drive 2, Singapore  117543}
\author{Siddarth K. Joshi}
\affiliation{Center~for~Quantum~Technologies, National~University~of~Singapore,
3 Science Drive 2, Singapore  117543}
\author{Alessandro~Cer\`{e}}
\affiliation{Center~for~Quantum~Technologies, National~University~of~Singapore,
3 Science Drive 2, Singapore  117543}
\author{Ad\'{a}n Cabello}
\affiliation{Departamento de F\'{i}sica Aplicada II, Universidad de Sevilla, E-41012, Sevilla, Spain}
\author{Christian~Kurtsiefer}
\affiliation{Center~for~Quantum~Technologies, National~University~of~Singapore,
3 Science Drive 2, Singapore  117543}
\affiliation{Department~of~Physics, National~University~of~Singapore,
2 Science Drive 3, Singapore  117542}

\begin{abstract}
We present an experimental test of the CHSH Bell inequality on photon pairs in
a maximally entangled state of polarization in which a value $S=2.82759 \pm
0.00051$ is observed. This value comes close to the Tsirelson bound of
$|S|\le2\sqrt{2}$, with $S-2\sqrt{2}=0.00084 \pm 0.00051$. It also violates
the bound $|S|\le2.82537$  introduced by Grinbaum by $4.3$
standard deviations. This violation allows us to exclude that quantum
mechanics is only an effective description of a more fundamental theory.
\end{abstract}

\maketitle

{\em Introduction} ---
Bell~\cite{Bell64} showed that the results of measurements on quantum systems cannot be explained by local theories, since they
violate certain inequalities among the correlations between the outcomes of measurements on two distant locations $A$ and $B$.
The simplest of these
Bell inequalities is the one by Clauser, Horne, Shimony, and Holt
(CHSH)~\cite{CHSH69},
which can be written as $|S|\le 2$, where the parameter $S$
is a combination of correlations $E(a_i, b_j)$ defined as
\begin{equation}
    S=E(a_0, b_0)-E(a_0, b_1)+E(a_1, b_0)+E(a_1, b_1)\,,
    \label{eq:bell_quantity}
\end{equation}
where $a_{0,1}$ and $b_{0,1}$ are measurement settings in $A$ and $B$, respectively, and each measurement has two possible outcomes, $+1$ or $-1$.
The correlations $E(a_i, b_j)$ are defined from the joint probabilities $P$
for outcomes $++, +-, -+$, and $--$  as
\begin{equation}\label{eq:correlations}
    E(a_i, b_j)=P(++)-P(+-) -P(-+)+P(--)\,.
    \label{eq:correlation_function}
\end{equation}
Tsirelson~\cite{Tsirelson80} showed that, according to quantum theory, $|S|$
has an upper bound of $2 \sqrt{2} \approx 2.82843$.
Popescu and Rohrlich~\cite{PR94} demonstrated that values up to $S=4$
were compatible with the no-signaling principle that prevents
superluminal communication.
This difference stimulated the search for principles singling out Tsirelson's
bound as part of the effort for understanding quantum theory from fundamental
principles.
So far, the following principles have been identified that enforce Tsirelson's
bound: information causality~\cite{PPKSWZ09}, macroscopic
locality~\cite{NW09}, and exclusivity~\cite{Cabello:2015us}.
Other principles, such as non-signaling~\cite{PR94} and nontriviality of communication complexity~\cite{VanDam00,BBLMTU06}, allow for higher values.

On the other hand, quantum theory introduces a cut between the observer and the observed
system~\cite{Bohr:1934}, but does not provide a definition of what is an
observer~\cite{Wheeler:1983tp}. To address
this problem, Grinbaum has recently tried to integrate the observer into
the theory~\cite{Grinbaum:2015tf}. For this purpose, he introduces a
mathematical framework
based on algebraic coding theory~\cite{MacWilliams:1983wu} that provides a
general model for communication, and enables an information-theoretic
definition of an observer.
This definition involves a limit on the complexity of the strings the observer
can store and handle. These strings contain all descriptions of states allowed
by quantum theory, but may also contain information not interpretable in terms
of preparations and measurements.
The language dynamics of these strings leads to a continuous model in the
critical regime that, when restricted to measurements on bipartite systems
in a three-dimensional Euclidean space, predicts that the violation of
the Bell CHSH inequality is upper bounded by 2.82537(2).
This prediction holds under the assumption that the number of strings
with the same complexity after uncomputable Kolmogorov reordering is 6
and some assumptions on the mappings between certain metric spaces
(see~\cite{Grinbaum:2015tf}, Sec.~V). It further uses the most precise
determination available of a critical exponent in three-dimensional Ising
conformal field theory~\cite{ElShowk:2014jf}.

The value predicted by Grinbaum is slightly smaller than the Tsirelson bound,
and is so far consistent with all the available experimental results~\cite{
freedman:72,
Aspect:1981ga,
kwiat95,
weihs98,
Tittel98,
kwiat13,
Nawareg:2013vc,
Christensen:2015,
rowe01,
monroe04,
matsukevich08,
hofmann:12,
ansmann09,
hanson12}.
Not being able to exceed Grinbaum's limit would support that quantum theory is
only an effective description of a more fundamental theory~\cite{Grinbaum:2015tf}, and would have a
deep impact in physics and quantum information processing.
This has important consequences for
cryptographic security~\cite{FFW11}, randomness
certification~\cite{PAMBMMOHLMM10}, characterization of physical properties in
device-independent scenarios~\cite{MY04,BNSVY13}, and certification of quantum
computation~\cite{RUV13}.

An interesting aspect of Grinbaum's work is the prediction that Tsirelson's
bound is {\em experimentally unreachable}, while quantum physics does not
impose such a limit. The model can thus be compared with direct observations
in nature.

From a more general perspective, an experimental search for  the maximal
violation of a Bell inequality \cite{Bell64} tests the principles that predict
Tsirelson's bound~\cite{PPKSWZ09,NW09,Cabello:2015us} as possible explanations
of all natural limits of correlations.

\begin{figure}
  \begin{center}
    \includegraphics[width=\columnwidth]{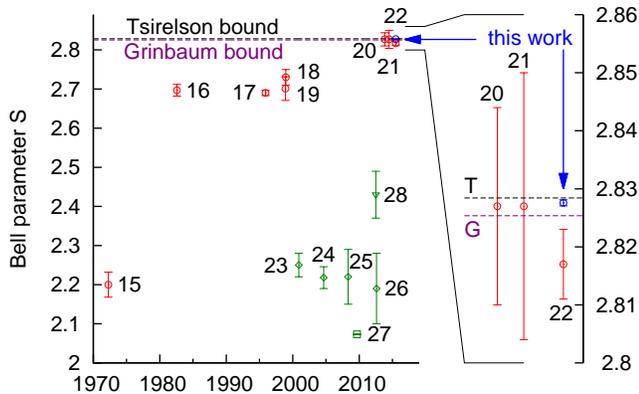}
  \end{center}
  \caption{\label{fig:vio}
    Selected experimental tests of the CHSH Bell inequality with results close
    to the Tsirelson (T) and Grinbaum (G)
    bounds in photonic systems (circles), atoms and ions (diamonds), Josephson
    junctions (square), and nitrogen-vacancy centers in diamond
    (triangle). Numbers represent the references.
  }
\end{figure}

{\em Prior work} --- The violation of Bell's inequality has been observed in many experiments with
exceedingly high statistical significance.
Many of these experiments are based on the generation of correlated photon pairs
using cascade decays in atoms~\cite{freedman:72,Aspect:1981ga},
or exploiting non-linear optical processes~\cite{kwiat95,weihs98,Tittel98,
Nawareg:2013vc,kwiat13}.
Other successful demonstrations were based on
internal degrees of freedom of ions~\cite{rowe01,monroe04,matsukevich08} and
neutral atoms~\cite{hofmann:12}, Josephson junctions~\cite{ansmann09}, and nitrogen-vacancy
centers in diamond~\cite{hanson12}.
Figure~\ref{fig:vio} summarizes the result obtained for the Bell parameter and
the corresponding uncertainty of several experimental tests.

While continuous experimental progress has made it possible to
approach Tsirelson's bound with decreasing uncertainty, predictions such as
Grinbaum's, which would imply a radical departure from standard
quantum theory, are compatible  with all existing results.

Here, we report on an experiment with entangled photon pairs that pushes the uncertainty in the Bell parameter by
another order of magnitude compared to previous experiments.

\begin{figure}
    \begin{center}
        \includegraphics[width=0.9\columnwidth]{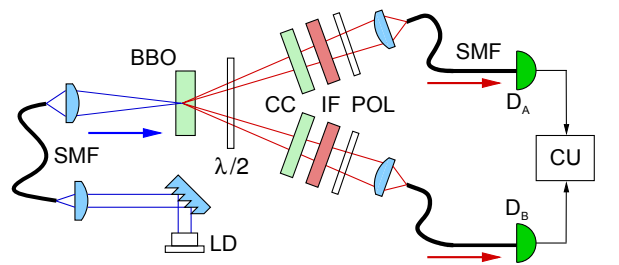}
        \caption{Schematic of the experimental set-up. Polarization
          correlations of entangled-photon pairs are measured by film
          polarizers (POL) placed in front of the collection optics. All
          photons are detected by silicon avalanche photodetectors D$_A$ and D$_B$, and registered in a coincidence unit (CU).}
        \label{fig:experimental_setup}
    \end{center}
\end{figure}

Our experiment follows the concept in~\cite{kwiat95} and is shown in
Fig.~\ref{fig:experimental_setup}. The output of a
grating-stabilized laser diode (LD, central wavelength 405\,nm) passes through
a single mode optical fiber (SMF) for spatial mode filtering, and is focused
to a beam waist of 80\,$\mu$m into a 2\,mm thick BBO crystal.

In the crystal, cut for type-II phase-matching,
spontaneous parametric down-conversion (SPDC) in a slightly non-collinear
configuration generates photon pairs.
Each down-converted pair consists of an ordinary and extraordinarily polarized photon, corresponding to
horizontal (H) and vertical (V) in our setup.
Two SMFs for 810\,nm define two spatial modes matched to the pump mode to optimize the collection~\cite{kurtsiefer:01}.
A half-wave plate ($\lambda$/2) and a pair of compensation
crystals (CC) take care of the temporal and transversal
walk-off~\cite{kwiat95}, and allow to adjust the phase between the two decay
components to obtain a singlet state
$\ket{\Psi^-}=1/\sqrt{2}\left(\ket{H}_A\ket{V}_B-\ket{V}_A\ket{H}_B\right)$.

Film polarizers (specified extinction ratio $10^4$) perform the basis choice
and polarization projection. Photons are detected by avalanche photo diodes
(APDs, quantum efficiency $\approx$40\%), and corresponding
detection events from the same pair identified by a coincidence unit (CU) if
they arrive within $\approx\pm$1.2\,ns of each other.

To arrive at a very clean singlet state, we carefully align the photon pair
collection to balance the two photon pair contributions $\ket{HV}$ and
$\ket{VH}$, and adjust their relative phase with the CC. Furthermore, we
minimize contributions from higher order parametric conversion processes
\cite{Wasilewski08} by restricting the pump power below 7\,mW, leading to
average detection rates
of $(4.84\pm0.20)\cdot10^{3}$\,s$^{-1}$ and
$(3.45\pm0.25)\cdot10^{3}$\,s$^{-1}$ at the two the detectors (uncorrected for
dark counts), resulting in an accidental coincidence rate of 0.020$\pm
0.017$\,s$^{-1}$.
The rate of coincidence events depends on the orientation of the polarizers,
as expected, and, in our measurements, ranges from a minimum of 26\,s$^{-1}$ to a maximum of
217\,s$^{-1}$. The detectors exhibit dark count rates of
91.7\,s$^{-1}$ and 106.2\,s$^{-1}$, respectively.

We test the quality of polarization entanglement by measuring the polarization
correlations in the
$\pm$\,45$^\circ$ linear polarization basis. With interference filters (IF) of 5\,nm bandwidth (FWHM) centered at 810\,nm,
we observe a visibility $V_{45}=99.9\pm0.1\%$. The visibility in the natural H/V basis of the type-II down-conversion process also reaches $V_{\rm HV}=99.9\pm0.1\%$. This indicates
a high quality of polarization entanglement; the uncertainties in the
visibilities are obtained from propagated Poissonian counting statistics.

Due to imperfections in the state generation and errors in the setting of the
polarizers, the setting $\theta=22.5^\circ$
may not yield the maximum possible violation.
In order to observe the largest possible violation, and get as close as possible to the Tsirelson bound,
we optimized the angular settings of the polarizers.

The optimization starts by setting $a\,=\,0^\circ$. This provides the initial reference axis and corresponds to $a_0$.
Rotating $b$ and recording the rate of coincidences, we identify the angles $b_0'$ and $b_1'$ that better match the expected correlation values. Setting $b=b_0'$, we repeat a similar procedure for $a$, obtaining $a_0'$ and $a_1'$.
This procedure converged to the resolution of the rotation motors (verified
repeatability/resolution 0.1$^\circ$).
For our experiment the optimal angles are $a_0'\,=\,1.9^\circ$, $b_0'\,=\,22.9^\circ$, $a_1'\,=\,46.8^\circ$, and $b_1'\,=\,67.7^\circ$.

Each of the correlations $E$ in (\ref{eq:correlations}) is estimated from
coincidence counts $N$ between $A$ and $B$,
\begin{equation}\label{eq:correlation}
    E=\frac{N_{++}-N_{+-}-N_{-+}+N_{--}}{N_{++}+N_{+-}+N_{-+}+N_{--}}\,.
    \label{eq:fair_sampling}
\end{equation}
For evaluating how close we can come with the test of the CHSH Bell inequality
to the Tsirelson
bound with a known uncertainty, we need to integrate for a sufficiently long
time to acquire the necessary counting statistics, assuming we have the
usual Poissonian statistics implied by the time invariance of our
experiment. We collect coincidence events for each of the 16 settings
required to evaluate $S$ for 1 minute, and then repeat again the whole
set. Within 312 such complete sets, we registered a total of 33,184,329 pair
events. As a result, we obtain in this experiment, via
Eqs.~(\ref{eq:bell_quantity}) and (\ref{eq:fair_sampling}), a value of
$S=2.82759\,\pm\,0.00051$, or a separation of $2\sqrt{2}-S=0.00084\pm0.00051$
from the Tsirelson bound.

The uncertainty we report on this quantity has several contributions. In the
following, we go through those we could identify.

Counting statistics -- The parametric down conversion process delivers
detection events randomly without any specific dynamics. Therefore, the
uncertainties in the coincidence events $N$ entering the correlation functions
$E$ via (\ref{eq:correlation}) show a Poissonian statistics.
The contribution from this, propagated through (\ref{eq:correlation}) and
(\ref{eq:bell_quantity}), is $\Delta S_P=4.9\cdot10^{-4}$.

Detector efficiency -- It is reasonable to assume that the quantum
efficiency of Silicon APDs remains stable over
the time necessary for each measurement of a correlation $E$, approximatively 10 minutes.
Single event rates detected for this experiment, approximatively
5000~s$^{-1}$, are low enough so that the response of the detector is,
effectively, linear. Thus, we do not assign any uncertainty in $S$ to any
efficiency drift in the detectors.

Detector dead time -- The passively quenched Silicon APDs we used have a dead
time of approximately 1.6~$\mu$s. Fluctuations in the total acquisition time
due to the dead time are proportional to the statistical fluctuations in count
rate, i.e., the square root of the number of single detection events.
Propagating this uncertainty to the calculated value of $S$, we obtain an
uncertainty $\Delta S_D=5.4\cdot10^{-7}$.

Timing uncertainty -- The counting intervals of 60~s are defined by a
hardware clock in a microcontroller, with a maximum time uncertainty of 100~ns.
This time jitter contributes an uncertainty $\Delta S_T=4.7\cdot10^{-11}$.
The temperature dependence of the reference clock is
also a source of timing uncertainty.
The maximum frequency drift of this clock we measured in a similar thermal
profile against a Rubidium-stabilized reference oscillator is in less than 0.1
ppm (part per million), leading to an uncertainty of 
$\Delta S_C=2.8\cdot10^{-9}$.

Angular position of polarizers -- From the angular uncertainty of 0.1 degrees
of the polarizer rotation stages, we estimate a contribution $\Delta
S_R=1.2\cdot10^{-4}$.

The resulting uncertainty quoted above is obtained via $\Delta S=(\Delta S_P^2+\Delta
S_D^2+\Delta S_T^2+\Delta S_R^2)^{1/2}$.
This analysis suggests that $\Delta S$ is
dominated by counting statistics, i.e., the total number of registered count
events.
Our experiment has certainly systematic uncertainties - for example, we do
observe an effective setting-dependent variation of the detection efficiency
due to small wedge errors in the film polarizers in front of the single mode
optical fiber collection optics on the order of a few percent. However, any
bias of this kind lowers the value of $S$ (attacks on detectors
excluded~\cite{Gerhardt11}, i.e., under the fair sampling assumption).

{\em Conclusion} --- The result of our experiment violates Grinbaum's bound by $4.3$ standard
deviations and constitutes the tightest experimental test of Tsirelson's bound
ever reported. Therefore, it shows no evidence in favor of the thesis that
quantum theory is only an effective version of a deeper theory and reinforces the thesis that quantum theory is fundamental and that the Tsirelson bound is a natural limit that can be reached. This conclusion strengthens the potential value of those principles that predict Tsirelson's bound~\cite{PPKSWZ09,NW09,Cabello:2015us} for explaining the natural limits of correlations in all scenarios.
The possibility of experimentally touching Tsirelson's bound as predicted by quantum theory also has important consequences for cryptographic security, since a necessary and sufficient condition for certifying probability distributions independent of the results of an eavesdropper in a device-independent scenario~\cite{ABGMPS07} is that the observed probabilities are exactly the ones corresponding to the Tsirelson bound~\cite{FFW11}.
It is also important for the certification of a variety of physical properties based solely on the assumption of non-signaling (i.e., without making assumptions on the initial state of the system or the inner working of the measurement devices). In this respect, the degree of violation of the CHSH Bell inequality can be used to certify the amount of randomness~\cite{PAMBMMOHLMM10}.
The higher the violation, the larger the amount of certified randomness. Reaching the Tsirelson bound can also be used to certify that the state being measured is a maximally entangled state and/or that the local measurements are of the type represented in quantum theory by anti-commutating operators~\cite{MY04,Popescu:1992gi}.
This can be adapted to practical methods to estimate the fidelity of the maximally entangled estates~\cite{BNSVY13}.
Finally, saturating the Tsirelson bound can be used to certify that a general quantum computation was actually performed~\cite{RUV13}.

We acknowledge support
by the FQXi large grant project ``The Nature of Information in Sequential Quantum Measurements'', the project ``Advanced Quantum Information'' (MINECO, Spain) with FEDER funds, and the
National Research Foundation \& Ministry of Education in Singapore, partly through the Academic Research Fund MOE2012-T3-1-009.

\bibliographystyle{apsrev4-1}
%

\end{document}